\documentclass[useAMS,usenatbib]{mn2e}
\usepackage{graphics}
\usepackage{psfig}

\def\simlt{\mathrel{\rlap{\lower 3pt\hbox{$\sim$}}\raise 2.0pt\hbox{$<$}}}
\def\simgt{\mathrel{\rlap{\lower 3pt\hbox{$\sim$}} \raise 2.0pt\hbox{$>$}}}

\def\gtsima{$\; \buildrel > \over \sim \;$}
\def\ltsima{$\; \buildrel < \over \sim \;$}
\def\gtrsim{\lower.5ex\hbox{\gtsima}}
\def\lesssim{\lower.5ex\hbox{\ltsima}}
\def\url#1{{\ttfamily\def\/{/\diskretionary{}{}{}}#1}}

\newcommand{\q}{\begin{equation}}
\newcommand{\qa}{\begin{eqnarray}}
\newcommand{\qs}{\begin{eqnarray*}}
\newcommand{\nq}{\end{equation}}
\newcommand{\nqa}{\end{eqnarray}}
\newcommand{\nqs}{\end{eqnarray*}}

\begin{document}

\title[Merger and Ring Galaxy Formation Rates]{Merger and Ring Galaxy Formation Rates at z$\le$2}
\author[E. D'Onghia, M. Mapelli \& B. Moore]
{Elena~ D'Onghia$^{1}$\thanks{Marie Curie Fellow; Email: elena@physik.unizh.ch}, Michela Mapelli$^{1}$ \& Ben Moore$^{1}$
\\
$^1$ Institute for Theoretical Physics, University of Z\"urich, Winterthurerstrasse 190, CH-8057, 
Z\"urich, Switzerland\\
}

\date{submitted to MNRAS}
\pagerange{\pageref{firstpage}--\pageref{lastpage}}\pubyear{200?}
\maketitle \vspace {7cm }

\label{firstpage}

\begin{abstract}
We compare the observed merger rate of galaxies over cosmic time and the frequency of collisional ring galaxies (CRGs), with analytic models and halo merger and collision rates from a large cosmological simulation. In the Lambda cold dark matter model (LCDM) model we find that the cosmic {\it merger fraction} does not evolve strongly between 0.2$\le$z$\le$2, implying that the observed decrease of the cosmic star formation rate since z$\sim 1$ might not be tied to a disappearing population of major mergers. 
Haloes hosting massive galaxies undergo on average $\sim 2$ mergers from z$\le 2$ up to present day, reflecting the late assembly time for the massive systems and the related downsizing problem. The cosmic {\it merger rate} declines with redshift: at the present time it is a factor of 10 lower than 
at z$\sim 2$, in reasonable agreement with the current available data.
The rate of CRG formation derived from the interactions between halo progenitors up to $z=2$ is found to be a good tracer of the cosmic merger rate. 
In the LCDM model the rate of CRGs as well as the merger rate do not scale as $(1+z)^m$, as suggested by previous models. Our predictions of cosmic merger and CRG rates may be applied to forthcoming surveys such as GOODS and zCOSMOS.

\end{abstract}
\begin{keywords}

galaxies: interactions - galaxies: peculiar - methods: {\it N}-body simulations - cosmology: theory
\end{keywords}

\section{Introduction}
In a hierarchical Universe, galaxy mergers are thought to play an important role during structure formation, particularly at higher redshifts.
Mergers may be also relevant to the growth of massive early-type galaxies (e.g. Toomre 1977; Barnes \& Hernquist 1996, Naab et al. 2006, Cox et al. 2006). Simulations of mergers involving gas-rich discs suggest that major mergers can trigger violent starbursts and transform discs into spheroidals. 
However the role of mergers in the galaxy assembly and star formation processes is still unclear. 
The observed correlation between galaxy morphology and color indicates that the star formation history of a galaxy 
is closely tied to its morphology evolution.
However, it has been recently recognized that there may be different timescales for the formation of stars in massive spheroidals.  
Thus, tracking the galaxy merger rate as a function of redshift can constrain the contribution of mergers to the formation of stars in spheroids. 

Despite their importance, it has proved challenging to measure the rate of galaxy mergers and its evolution with cosmic time.
Many theoretical and observational attempts have attempted to reconstruct the history of the galaxy interaction rate (Toomre 1977; Zepf \& Koo 1989; Carlberg 1990a, 1990b; Carlberg, Pritchet \&{} Infante 1994; Burkey et al. 1994; Yee \& Ellington 1995; Neuschaefer et al. 1995; Woods, Fahlman, \&{} Richer 1995; Patton et al. 1997; Le F\`evre et al. 2000; Conselice, Bershady \& Jangren  2000a; Conselice, Bershady \& Gallagher 2000b; Bershady, Jangren \&  Conselice 2000; Conselice 2003; Conselice et al. 2003, 2004; Cassata et al. 2005; Conselice 2006; Bridge et al. 2007; Jogee et al. 2007; 2008; 
see Conselice 2007 for a short review).

Various models (Toomre 1977; Carlberg 1990a, 1990b) suggest that the galaxy merger rate per unit volume $\dot{n}$ increases with the redshift $z$ as:
\begin{equation}\label{eq:eq1}
\dot{n}\propto{}(1+z)^m,
\end{equation}
The theoretical approach proposed by Carlberg (1990a, 1990b), based on the Press$-$Schechter formalism (Press \& Schechter 1974), predicts that the value of the coefficient $m$ depends on the present-day matter density parameter $\Omega{}_{\rm M}$:
\begin{equation}\label{eq:eq2}
m\sim{}4.51\times{}\Omega{}_M^{0.42}.
\end{equation}

 A number of studies have used a sample of haloes or subhaloes using N-body simulations 
to estimate merger rates as a function of redshift. 
Governato et al. (1999) studied major mergers of galaxy-sized haloes in open CDM universe. 
Gottloeber et al. (2001) studied the environmental dependence of major merger
rates as a function of redshift in the concordance CDM models. 
More recently, the Millennium simulation (Springel 2005)  has been used to construct merger trees and to quantify the 
merger rates of haloes (Fakhouri \& Ma 2008). The authors find that the average merger rate per halo depends 
weakly on the halo mass and that 
the halo merger rate evolves with $z$ as
$(1+z)^{m}$  with $m$ in the range 2 to 2.3.  Guo \& White (2008) found similar results using the Millennium simulation galaxy catalog: the halo merger rates depend
on redshift and only weakly on the halo mass. Murali et al. (2002) studied the relative contributions
of merging and smooth accretion to the rate at which large galaxies gain mass using  cosmological cubes simulated with N-body
and SPH techniques. Maller et al. (2006) estimated the merger rates of galaxies identified within subhaloes
using cosmological  hydrodynamic simulations.
Additional studies used semi-analytic models to infer the merger fraction of haloes (e.g. Khochfar \& Burkert 2001;
Benson et al. 2005). Most of the models predict that the major merger rate of galaxy-sized dark matter haloes rises
rapidly with redshift. 

The number of close pairs is often used as an observational tracer of the galaxy merger rate. However
observational studies suggest of the number  of close companions evolve less, over the cosmic time, than
inferred from theoretical studies of dark halo merging.  
Barrier et al. (2006) estimated the major merger rates of subhaloes using analytical models plus numerical N-body
simulations and studied the connection of the merger rates to the observed number of close pairs. 
The authors use the halo-occupation-distribution (HOD) modelling of galaxy clustering
to explain the little evolution in redshift of the observed close-pair count. They find that the discrepancy is due to
the additional processes occurring during merger of subhaloes in a common parent halo, which may be
ignored in the halo merger rates.

Due to the difficulty of directly observing the merger features, measurement of the redshift evolution of the fraction of galaxy pairs has traditionally been taken, which can be parametrized as $\propto{}(1+z)^k$. Most  studies (Zepf \& Koo 1989; Carlberg et al. 1994; Burkey et al. 1994; Yee \& Ellington 1995; Woods et al. 1995; Patton et al. 1997; Kampczyk 2007) derive a value of $k{}\sim{}2-4$, while Neuschaefer et al. (1995) find $k\sim{}0$, due to a different estimate of the non-physical galaxy pairs. The main difficulty of this method is that the conversion from $k$ to $m$ is unclear. It is also 
difficult to disentangle projection pairs from
true physical pairs when using photometric redshifts alone. Additionally, 
not all galaxies in a physical pair will merge, as the galaxies
may be unbound.

From an observational point of view, it is difficult to estimate whether $\dot{n}$ depends on the 
redshift and to determine the correct value of $m$. A direct measurement of merger features in distant galaxies is difficult, as tidal tails and distortions  generally have low surface brightness (see Mihos 1995; Hibbard \& Vacca 1997).

A powerful method for measuring the galaxy merger rate is to count the incidence of strongly disturbed galaxies (with strong asymmetries, double nuclei or prominent
tidal tails), the so-called CAS method (Conselice 2003). There
are several methods to quantify the frequency of strongly distorted
galaxies: visual classification,  quantitative measures of asymmetries
such as the CAS system (Conselice 2003) and the Gini-M20 system (Lotz et
al 2006). 

All asymmetries suffer from surface brightness (SB) dimming, but the
outer low SB features suffer more strongly from it.  CAS misses the
latter features, but visual classification captures many of these. 
Simulations (Conselice 2006)
as well as empirical studies (Jogee et al. 2007, 2008) show that visual
classifications capture a larger fraction of strongly distorted  galaxies
than the CAS merger criteria, as the eye is sensitive to asymmetries
over a larger dynamic range.

The CAS method, applied to the {\it Hubble Deep Field} ({\it HDF}), 
provides an estimate of $m$ ranging from 4 to 6 (Conselice et al. 2003; but $m\sim{}2-4$ in the re-analysis of these data by Conselice 2006). Based on the results of the CAS analysis, Conselice (2006) suggests that equation (\ref{eq:eq1}) is inaccurate for some galaxy types (especially for small galaxies) and for high redshifts ($z\gtrsim{}1-2$).
The CAS method provides a more straight-forward estimate of merger rate than the incidence of close pairs of galaxies. However, 
positioning a galaxy in the CAS plane is sometimes problematic. Furthermore, the connection between high asymmetry/lumpiness's and merger history 
implies a number of assumptions. 

For these reasons, Lavery et al. (2004; hereafter L04) proposed to use ring galaxies as a more direct tracer of galaxy mergers.
In fact, a high fraction of ring galaxies [$\approx{}60$ per cent, Few \& Madore (1986)], 
called 'P-type ring galaxies', are thought to have collisional origin. Recent $N$-body simulations (Mapelli et al. 2008a, 2008b, and references therein) show that the ring phase is quite short-lived: it lasts only for $\lesssim{}500$ Myr after the galaxy collision. Moreover, ring galaxies are easier to identify than other interaction signatures (e.g. tidal tails; see Mihos 1995; Hibbard \& Vacca 1997).
Thus, the number of collisional ring galaxies (hereafter CRGs) may be a straight-forward tracer of the galaxy interaction rate.

Unfortunately, most  ring galaxies with measured distances are relatively nearby [$z\lesssim{}0.1$; see e.g. the sample of 68 ring galaxies in  Few \& Madore (1986)]. 

L04 analyze 162 Wide Field Photo Camera 2 (WFPC2) fields, obtained from the {\it Hubble Space Telescope} ({\it HST}) Archives, in order to identify distant CRGs. They find 25 CRGs in their images. From this sample, L04 derive a value of the merger rate $m\sim{}5$. However, their estimate is affected by large uncertainties, as  they have redshift measurements only for 6 of their 25 CRGs. For the remaining 19 galaxies, they derive an 'estimated redshift', by assuming that  CRGs have similar visual magnitude. 
Recently, Elmegreen \& Elmegreen (2006, hereafter E06) analyzed other 24 CRGs in the GEMS and GOODS fields. 
For these galaxies redshift measurements are available.

In this paper we present an attempt to derive the merger and the CRG formation rate from cosmological simulations. 

The simulations can estimate the rate of minor and major mergers in progenitors of the present-day galaxy haloes and 
establish whether the merger rates and CRG formation rates are related. We calculate the evolution of the merger and CRG formation rate up to redshift 2.
We use a cosmological cube a factor of three larger and higher numerical resolution than adopted in the SPH simulations of Maller et al. (2006) and 
our numerical resolution is higher of almost a factor of ten than the Millennium run used in Fakhouri \& Ma 2008).
Finally this study  compares results from numerical simulations with the available data and gives predictions for future observations.
 In Section~2 we present details of the numerical simulation and analysis procedure.
Section~3 discusses our main results, 
 while Section~4 summarizes our
conclusions and implications.

\section{Numerical Methods}
\subsection{Simulations}
We analyze a cosmological $N$-body simulation of the Lambda cold dark matter (LCDM) cosmogony, with cosmological parameters chosen to match the 3-yr Wilkinson Microwave Anisotropy Probe (WMAP3)
constraints (Spergel et al. 2007). These are characterized by the
present-day matter density parameter, $\Omega_{\rm {M}}=0.238$, a cosmological
constant contribution, $\Omega_{\Lambda}$=0.762, and a Hubble
parameter $h=0.73$ ($H_0=100\, h$ km s$^{-1}$ Mpc$^{-1}$). The mass
perturbation spectrum has a spectral index of $n=0.951$, and is
normalized by the linear rms fluctuation on $8$ Mpc/$h$ radius
spheres, $\sigma_8=0.75$.

We follow the evolution of $600^3$ particles of mass $m_{\rm{dm}}=
8.67 \times 10^{7}  \, h^{-1} \,M_{\odot}$ in a box of 90 Mpc (or 65.7 Mpc/h)
(comoving) on a side, by using the $N$-body code PKDGRAV 
(Stadel 2001). Gravitational interactions between pairs of particles
are softened with a fixed comoving softening 
length of 1.16 kpc.

\subsection{Halo Identification at z=0}
Non-linear structures at $z=0$ are identified using the classic
friends-of-friends (FOF) algorithm with a linking length equal to
$0.2$ times the mean comoving interparticle separation.  For each FOF halo we
identify the most bound particle and adopt its position as the halo
centre. Using this centre, we compute the `virial radius' of each
halo, $r_{\rm vir}$, defined as the radius of a sphere of overdensity
$\Delta(z=0)=94$ (relative to the critical density for
closure)\footnote{The virial overdensity in a flat universe may be
computed using the fitting formula proposed by Bryan \& Norman (1998):
$\Delta (z)= 18 \pi^2 + 82\, f(z)-39\, f(z)^2$; with
$f(z)=\frac{\Omega_0(1+z)^3}{\Omega_0(1+z)^3+\Omega_{\Lambda}}-1$}. Quantities
measured within $r_{\rm vir}$ will be referred to as `virial', for
short. We select for our analysis all haloes with masses in the range
$M_{\rm vir}=5 \times 10^{12}$ to $10^{14} \, h^{-1}$ M$_{\odot}$. The resulting haloes 
have $N_{\rm vir}$ between $56,000$ and $1,100,000$
particles within the virial radius.

\subsection{Merger tree Construction}
For each of the haloes in our $z=0$ sample we have constructed
a merger tree over the period $0<z<2$, based  on FOF haloes. We use the FOF merger tree
to define and quantify the accretion and merger rates of haloes. Other authors have previously
used this technique or used substructure directly. We prefer to avoid the use of substructure
since at a fixed mass resolution, smaller haloes are less resolved and have less substructure
than larger haloes due to the classic overmerging problem (Moore et al 1996).

We consider a FOF halo identified at $z>0$ to be 'progenitors' of a $z=0$ system if at least 50 per cent
of its particles are found within the latter. 
Using this definition we can identify, at all times, the list
of progenitors of a given $z=0$ halo and track their properties
through time.
In the tree to identify halo 
mergers, we denote a halo as a major merger remnant if at some time during 
$0<$z$<2$ its major progenitor was classified as a single group in one output but two separate 
groups with a mass ratio $\le 4 : 1$ in the preceding output (see D'Onghia \& Burkert 2004).
As already noted in previous works (e.g. Gottloeber et al. 2001, Berrier et al. 2006, Fakhouri \& Ma
2008), the merger tree can result in  fragmentation events, in which particles
of a progenitor halo end up in two distinct halo descendants.
This spurious fragmentation is an artifact of the FOF halo identification scheme. During 
the initial merge phase the halo finder associates
and dissociates particles inside and outside the bound region. 
We noted that the fragmentation event of a halo progenitor lasts for
less than 0.5 Gyr. After this time usually the progenitor merges again.
If this fragmentation is not properly treated, the risk is to count twice the same merger event for 
the current halo. We avoid this risk by assuming that the merger happens the first time
the progenitor halo is considered part of the descendant. Our method to handle  fragmentation events
is similar to the `stitching' method assumed from Fakhouri \& Ma (2008).
Each FOF halo at $z>0$ in the tree catalog is assigned a mass
counting the number of particles associated with the FOF group, rather
than a mass defined with the overdensity criterion.
Halo progenitors of interest contain at least $250$ particles.

An additional problem with the merger tree is that the mass of a  descendant halo
is not exactly the sum of all the progenitors, but it may count inside the halo  a diffuse 
mass component which is not resolved in subhaloes due to limited numerical resolution. 
The diffuse mass component may dominate the merger events at high redshift,
when it is more difficult to resolve progenitors, owed the limited numerical resolution. 
However our current numerical
resolution is almost a factor of ten higher than the Millennium run used from Fakhouri \& Ma (2008) 
and of the SPH simulation of Maller et al. (2006)
and allows us to resolve progenitors and relative mergers and  
guarantees that we do not miss any relevant major merger event and 
 interaction of haloes progenitors back in time.

Previous studies built the halo merger trees by connecting subhaloes instead of FOF haloes across the snapshot outputs
(Gottloeber et al. 2001, Maller et al. 2006, Berrier et al. 2006).
In these papers the procedure to define descendants and progenitors
is similar to that one assumed for FOF trees: a subhalo at a given redshift is assumed  to be descendant 
of a progenitor subhalo defined at higher redshift if it contains a fixed percentage
of particles. In particular Berrier at al. (2006) uses a hybrid N-body simulation plus analytic substructure model to predict 
the number of pairs, which is the quantity often used to infer the observed galaxy merger rate. Since
the observed number of close companions rises with redshift slower than the halo major merger rates predicted from
previous simulations, they assume a halo-occupation model, in order 
to match the observed number of close companions inferred
from current data.  
Indeed, our algorithm is based on a FOF halo tree and does not include subhaloes. However, a merger tree
constructed on subhaloes inside a parent halo is mainly required to analyze galaxy-cluster sized haloes 
and massive galaxy-group sized haloes. The cosmological cube considered here contains only one cluster-sized halo.
 The galaxy-sized haloes and galaxy-group sized haloes 
defined in our sample are located in low density environments  and are compared to 
galaxies which are not located in clusters, but mostly  in the field.

It worth noting that when merger trees are based on subhaloes it is 
difficult to estimate the mass of a subhalo located within a larger halo.
Tidal stripping occurring when a subhalo enters into larger halo can reduce significantly the mass
of the subhalo before it approaches the pericentric distance, with implications for the inferred
mass ratio of the merging events. 
Since our merger trees are based on FOF haloes and do not consider subhaloes, our analysis does not suffer of this problem.


\subsection{Ring Galaxy Formation Criteria}

Since our simulations are based on  dark matter only, we cannot directly trace  the formation of CRGs.
However, we can estimate the rate of CRG formation  from our models  adopting the following criteria. 
At each redshift we select a sample of progenitor haloes in the range of mass of a few $10^{11}\,{}M_{\odot}$
to $5\times{}10^{12}\,{}M_{\odot}$ (corresponding to the typical masses of observed CRGs) and we consider the encounters  between 
them. 
We consider a list of 8 progenitors ordered by decreasing mass and 
we examine the collisions between the most massive and the second most massive progenitor 
and all the possible
combinations amongst progenitors along the list. 

We assume that a CRG is formed whenever two haloes undergo an encounter in which: 
\begin{itemize}
\item[ i)] the mass ratio  between  the bullet (hereafter 'intruder' halo) 
and the most massive progenitor (hereafter 'target' halo)  is $\ge 1:10$;

\item[ ii)] the pericentric distance $p$  is less than $\sim{}$15 per cent of the expected disc radius of the target galaxy (i.e. $ p \le 4$ kpc, Lynds \& Toomre 1976).
\end{itemize}
The first constraint comes both from  observations of  nearby CRGs whose intruder is known, and from  numerical  simulations showing that the ring 
is hard to form when the mass ratio between the intruder
and the target galaxy is $\le 1:10$ 
(Hernquist \& Weil 1993; Mihos \& Hernquist 1994; Horellou \& Combes 2001; Mapelli et al. 2008a, 2008b). 

The second condition assumes results from numerical simulations showing that circular rings 
form only when the impact parameter is small.
The larger the impact parameters, the more asymmetric is the resulting ring
(Hernquist \& Weil 1993; Mihos \& Hernquist 1994; Horellou \& Combes 2001; Mapelli et al. 2008a, 2008b).

Finally, we do not put any limitation on the inclination angle $\theta{}$ (between the disc axis of the target and the velocity of the intruder). Lynds \& Toomre (1976) indicated that relatively symmetric ring galaxies can form at least for $\theta{}\le{}45^\circ{}$. Recent simulations (Ghosh \& Mapelli 2008) show that regular (although warped) rings form also for $\theta{}>60^\circ{}$. So, we can reasonably assume that ring galaxies can  form also for high values of $\theta{}$. Thus, our estimate of the CRG formation rate represents an upper limit and should be rescaled by an unknown factor $1-\cos(\theta{})\ge{} 0.5$.

\section{Results}

\subsection{Merger Fraction Evolution up to Redshift 2}

To predict the merger fraction as a function of redshift using observations of galaxy pairs is a difficult task, due to the  complexity of establishing the time 
over which the merger occurred. Estimates of merger fractions using galaxy pairs  come from
Patton et al. (2000, 2002), Le F\`evre et al. (2000) and Lin et al. (2004) for $z<1$. However, 
owing to projection effects, some pairs might not be a physically bound system and merging may last for a long time or may never occur. 

On the other hand,
in simulations we assume that a  merger  occurs when the smaller halo enters into the virial radius of the host halo,
whereas the distance between the observed pairs is generally smaller than the virial radius of the progenitor. 
In fact, the baryonic component of the two galaxies is within $\approx{}20$ per cent of the virial radius.

A comparison of the merger rates extracted from the simulations for dark haloes 
 to galaxy merger rates has some limitations. A direct comparison implies a conversion between halo mass
and galaxy mass/light.  Many models usually adopted to this conversion,
e.g., see van den Bosch et al. (2007) and reference therein,   have
a strong dependence upon halo (or galaxy) mass. These models imply 
that a merger between a dark matter halo and another halo
of one-tenth its mass may not be equivalent to a merger between a
galaxy and another one of one-tenth its mass. 
Over the mass range $10^{11}M_\odot$ to $10^{12}M_\odot$, the variation
in M/L ratio with halo mass is approximately a constant in the HOD models, so
this approximation is good. For larger mass haloes we should compress our merger
mass scale by ~30\%, but this depepence is model based and would introduce an error
smaller than the observations currently suffer from. Ideally we would
use hydrodynamical simulations that accurately resolve the star-formation and
luminosity evolution of galaxies. This is a few years away before becoming 
achievable computationally.


The timescale  over which the merger between a pair 
of galaxies is deemed visible is estimated as a fraction of  the crossing time $t_{\rm{cr}}$ of the infalling halo into the host halo.
The crossing time for realization of  1:1 merger simulations is $\approx 1$ Gyr. 
Simulations of 1:1-1:2-1:3 mergers show 
that the timescale  over which merger features are visible is  $\approx 0.5$ Gyr for systems with different orbital inclinations (Conselice 2006).

We therefore define the halo merger fraction $f_{\rm{merg}}$ as 
\begin{equation}
 f_{\rm{merg}}(t,M)=\frac{N_{\rm{merg}}}{N_{\rm{tot}}}\frac{t_{\rm{cr}}}{\Delta t},
\end{equation}
where  $N_{\rm{tot}}$ and $N_{\rm{merg}}$  are the total number of haloes and the number of haloes undergoing a (minor or major) merger at 
a given time interval $\Delta{}t$ and for a given mass $M$, respectively.  $t_{\rm cr}/\Delta{}t$ is the fraction of the time interval over which the merger signature is visible.
In the time interval between z=0 and z=0.3 the time is longer than the merger time so we did not account for the time delay  $t_{\rm cr}/\Delta{}t$.

\begin{table*}
\caption{\hspace{3cm} Minor and Major Mergers.}
\begin{tabular}{cccccc}
\hline
\vspace{0.1cm}
$z$ &  Mergers$^{\rm a}$ & Major Mergers$^{\rm a}$ &     Mergers$^{\rm b}$         &    Major Mergers$^{\rm b}$ & N$_{\rm{tot}}$ Haloes  \\
    &                    &                         & [M$\sim 10^{12}\,{}M_{\odot}$]  & [M$\sim 10^{12}\,{}M_{\odot}$]
&  \\  
\hline
0.0  & 571 & 11  & $-$ & $-$ &  - \\
0.1  & 837 & 19  &  3  &  3  & 756  \\
0.3  & 649 & 30  &  11 &  7  & 1565  \\
0.5  & 655 & 17  &  30 &  8  & 2232  \\
0.7  & 706 & 43  &  65 &  26 & 2844  \\
0.9  & 496 & 4   &  74 &  3  & 3436  \\
1.1  & 424 & 21  &  91 &  12 & 3852  \\
1.3  & 326 & 6   &  87 &  2  & 4240  \\
1.4  & 254 & 10  &  91 &  8  & 4493  \\
1.6  & 267 & 15  &  88 &  7  & 4669  \\
1.7  & 302 & 14  &  82 &  7  & 4813  \\
1.9  & 210 & 14  &  69 &  6  & 5020  \\
\hline
\end{tabular}
\begin{flushleft}
{\footnotesize $^{\rm a}$ Total Mergers(minor+major) and major mergers from simulations for haloes at present day in the range of mass between $5\times{}10^{12}$ and 10$^{14}\,{}M_{\odot}$.\\
\footnotesize $^{\rm b}$ Total Mergers(minor+major) and major mergers from simulations for progenitors with fixed mass between $9\times{}10^{11}$ and $3\times{}10^{12}\,{}M_{\odot}$ at any redshift.
}
\end{flushleft}
\label{tab_1}
\end{table*}

\begin{figure}
\center{{
\psfig{figure=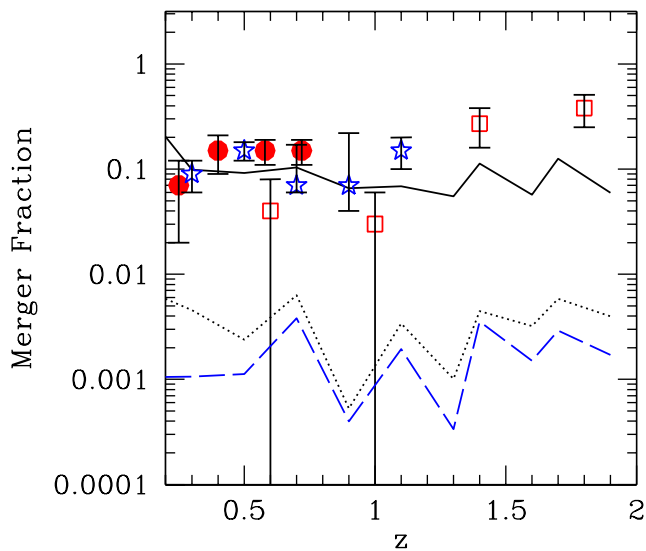,height=8cm}
}}
\caption{\label{fig:frac}Total merger fraction (solid line) and major merger fraction (dotted line) of dark haloes with mass between $5\times{}10^{12}$ and $10^{14}\,{} M_{\odot}$ at present day, 
as a function of redshift, derived from cosmological simulations. 
The long-dashed line shows the major merger fraction of Milky-Way sized haloes at any time.
Symbols represent empirical results on the merger fraction or fraction of
 strongly distorted galaxies identified via different methods, ranging from
 visual classification to  automated methods:  Jogee et al.  (2007, 2008,
 filled  circles), Le F\`evre (2000, filled squares),   Conselice (2003,
 open squares) and Lotz et al. (2006, stars). Note that a stellar mass cut off was applied in Jogee data points ($M_\ast{}>2.5\times{}10^{10}M_\odot{}$)
and in Conselice 2007 ($M_\ast{}>1\times{}10^{10}M_\odot{}$). 
}
\end{figure}

Table~1 lists the number of major and minor mergers and the total number of haloes derived in simulations.
In Figure \ref{fig:frac} major merger fractions (the dotted line) and total (i.e. minor+major) merger fractions (solid line) derived from simulations  
are plotted as a function of redshift when assuming $t_{\rm cr}=$ 1 Gyr. Our models predict a merger fraction
that does not evolve significantly with redshift between 0.2$\le$z$\le$2. This implies that the  observed decrease
of the cosmic star formation rate since z$\sim$1 (e.g. Lilly et al. 1996; Madau et al. 1996) 
is not tied to a disappearing population of major mergers, and seems to be in agreement with recent 
new results by Jogee et al. (2007; 2008) for $z\sim{}0.2-0.8$, and by Wolf et   al. (2005) and Bell et al. (2005) at $z\sim{}0.7$. 
 Table~1 also lists the number of total mergers (minor+major) and major mergers for Milky Way sized progenitors
identified at any time, with typical mass of $\sim 10^{12}\,{}M_{\odot}$.

 In Figure  \ref{fig:frac} we compare the predictions of our models with data of the merger fraction,
 or fraction of  strongly distorted galaxies, identified via different methods
 and based on different surveys  : Jogee et al. (2007, 2008, filled  circles),
 Le F\`evre (2000, filled squares),   Conselice (2003, open squares) and Lotz et
 al (2006, stars).   The results of   Jogee et al. (2007, 2008) refer to the
 the fraction of strongly distorted interacting/merging  massive (with stellar mass $M_\ast{}> 2.5\times{}10^{10}\,{}M_\odot{}$)
 galaxies over  $z\sim{}0.24$ to 0.80, identified from the GEMS survey  (Rix et al. 2004)
 using both the CAS system  and  an independent visual classification system,
 specifically  designed to separate interacting galaxies with externally-triggered
 asymmetries from non-interacting galaxies with small-scale internally-triggered
 asymmetries. 

The presence of wiggles in the theoretical estimates is due to the uncertainty of using the merger tree
to identifying mergers between progenitors at each time.  The major merger fraction of Milky-Way sized haloes identified at any time is displayed with the long dashed line. 
We note that the predicted fraction of major mergers and minor mergers is almost constant from z=2 up to present day for Milky-Way sized haloes identified at any redshift
and for all the haloes of our sample, regardless the mass.  
Furthermore, the agreement between merger fractions predicted from cosmological simulations (based on the halo merger history) and 
the observed merger fractions based on galaxy CAS morphologies is encouraging for the LCDM model.

\subsection{Merger Rates}
The merger fraction is related to the  merger rate per unit volume $\it{R}$, defined within a time interval and mass range, by the following expression.

\begin{equation}
\it{R} (t,\rm{M})=f_{\rm{merg}}\,{}\tau_m^{-1}\,{}n_m,
\end{equation}

where $\tau_m$ is the timescale for a merger to occur and is defined as $\approx t_{cr}$ 
and $n_m$ is the physical
density of the haloes undergoing  a merger (minor or major) within a given mass range and at a given time.
The physical density of merging haloes is derived by dividing the total number of haloes which are merging 
$N_{\rm{merg}}$, listed in Table~1,
by the physical volume occupied by all the considered haloes.

\begin{figure}
\center{{
\psfig{figure=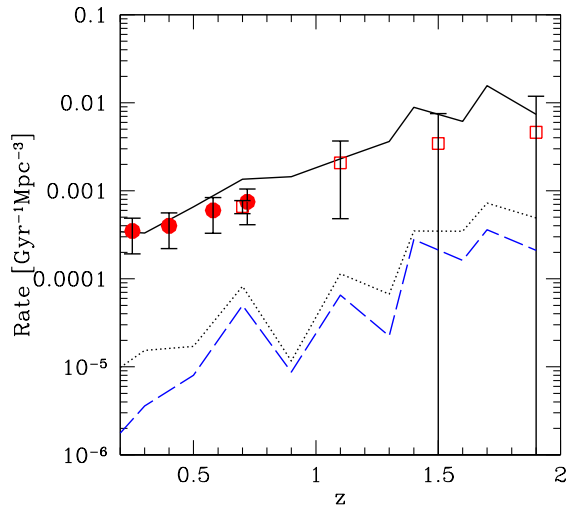,height=8cm}
}}
\caption{\label{fig:ratejog} 
Total merger rate (major+minor) (solid line) and major merger rate only (dotted line) of dark matter haloes, 
per unit volume (comoving), as a function of redshift, derived from cosmological  simulations. 
The rate per unit volume of Milky Way sized haloes identified at any time is plotted for comparison (long-dashed line, blue on the web). 
Filled circles: rate of mergers and interactions, with mass ratios in the
  range 1:1 to 1:10 , derived by  Jogee et al.  (2007, 2008) from the
  GEMS data. Open squares: merger rate from the HDF
  from Conselice (2003).}

\end{figure}

Figure~\ref{fig:ratejog} plots the major merger rate (dotted line) and the total merger rate 
(minor+major) (solid line) per unit volume, as a function of redshift, as derived from cosmological simulations. 
For comparison, Figure~\ref{fig:ratejog}  also shows the merger rate inferred from the GEMS sample (Jogee, private communication;
marked with filled circles) and the analysis of Conselice et al. (2007) based on the {\it Hubble Ultra Deep Field} ({\it HUDF}, open squares).
For completeness, the plot displays the major merger rate per unit volume inferred in cosmological models for progenitors in the 
range of mass of the Milky way ($\sim 10^{12}\,{} M_{\odot}$), identified  at any redshift (long-dashed line).
A large fraction of the progenitors are in this range of mass at higher redshift, explaining why these systems show a similar trend in the merger rate as the major 
mergers of all the haloes. The merger rate of Milky Way sized haloes (long-dashed line) decouples from the total major merger rate (dotted line) only at  $z\lesssim{}0.5$,  where the former drops, while the latter decreases more gently. 
This is due to the fact that mergers occurring at late time mainly involve the assembly of larger mass systems.

Note that the decrease of the volume-averaged merger rate at late times is a result of the decrease
of the progenitor number density at lower redshift.
Our predictions are in good agreement with the observations, despite the large uncertainties
in the available data. It worth noting that our estimates based on dark matter haloes  are in agreement with the estimates inferred from Maller et al. (2006) who used
N-body simulations plus hydrodynamics with a merger tree based on subhaloes to better trace the close pairs of galaxies.
It is encouraging to note that estimates based on dark haloes and SPH simulations do agree without invoking any additional treatments of baryons in galaxies 
with the halo-occupation models and semi-analytic models.

Summing over all mergers for massive haloes ($>5\times{}10^{12}\,{}M_{\odot}$), 
we find that the average number of mergers a halo experienced since $z\sim 2$ is $N_m \sim 2$ in agreement with
estimates of Conselice et al. (2007).

\subsection{Ring Galaxy Formation Rate}

\begin{figure}
\center{{
\psfig{figure=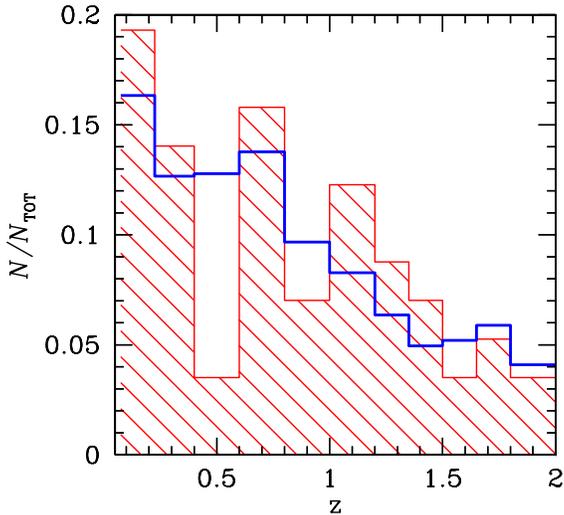,height=8cm}
}}
\caption{\label{fig:fig4} CRG formation rate (i.e. the number of newly formed CRGs in the entire simulation per redshift interval, lightly hatched histogram, red on the web) 
as a function of redshift, compared with 
the merger rate obtained considering all the haloes in the range of mass between $5\times{}10^{12}$ $M_{\odot}$ and $10^{14}$ $M_{\odot}$ at $z=0$ (open histogram, blue on the web; see Table~1 and Fig.~2).
The histograms are normalized to total number of CRGs and mergers, respectively.
}
\end{figure}

First we check the correspondence between the CRG formation rate and the halo merger rate within our cosmological simulation. For CRGs we consider only halo progenitors with mass between a few $10^{11}$ and $5\times{}10^{12}$ $M_{\odot}$ at any time, which correspond to the observed masses of ring galaxies. Figure~\ref{fig:fig4} shows the CRG formation rate (lightly hatched histogram, red on the web), compared with the total (minor+major) merger rate of all the haloes with present day masses between $5\times{}10^{12}$ $M_{\odot}$ and $10^{14}\,{}M_{\odot}$ (open histogram, blue on the web). 
Both the CRG formation rate and the merger rate increase at lower redshifts.
The null hypothesis probability that ring galaxies and mergers are drawn from the same 
distribution from redshift 0.2 up to z=2 is $\sim{}0.11$, corresponding to a non-reduced $\chi{}^2=17.1$ (for 11 data points, 0 parameters, 
and assuming Poissonian errors).
Thus, we conclude that the CRG formation rate can be considered a tracer of the merger rate.
Since our results rely upon the distribution
    of merging orbits we checked the consistency with previous works.
We computed the fraction of merging progenitors with different pericentric distances 
using the sample we adopted for the CRG calculations.  By tracking
accretion events we predict the pericentric distance of the merging haloes. This distribution
does not disagree with that  shown in Khochfar \& Burkert (2006) and 
is similar to that found in simulations that resolve substructures in CDM haloes (e.g. Ghigna et al 1998).

Finally, we remark that our simulations follow the merger history and dynamics of dark haloes and 
neglect processes involving the baryonic physics. In particular, the formation of galaxy discs,
which is crucial for the formation of CRGs, cannot be modeled here. Also the time evolution for 
the late and early type number fraction is not accounted. 

\subsection{Comparison with Observations}

The available data of moderately high redshift CRGs consist of a sample of 25 ring galaxies ($0.2\le{}z\le{}1$), observed with {\it HST} (LO4), and a more recent sample of 24 ring galaxies ($0.07\le{}z\le{}1.5$) found in the GEMS and GOODS fields (E06). 
Table~2 reports the number of CRGs observed in these two samples, as well as the number derived from our simulations.

\begin{table}
\begin{center}
\caption{Number of simulated and observed CRGs.}
\begin{tabular}{cccc}
\hline
\vspace{0.1cm}
$z$ &  simulated CRGs$^{\rm a}$ & L04$^{\rm b}$ &  E06\\
\hline
0.1  & 11 & $-$ &  1 \\
0.3  &  8 & 2   &  2 \\
0.5  &  2 & 7   &  4 \\
0.7  &  9 & 11  &  4 \\
0.9  &  4 & 5   &  8 \\
1.1  &  7 & $-$ &  4 \\
1.3  &  5 & $-$ & $-$ \\
1.4  &  4 & $-$ &  1 \\
1.6  &  2 & $-$ & $-$ \\
1.7  &  3 & $-$ & $-$ \\
1.9  &  2 & $-$ & $-$ \\
\hline
\end{tabular}
\end{center}
\begin{flushleft}
{\footnotesize $^{\rm a}$ CRGs from our simulation (considering haloes with mass between $\sim{}10^{11}$ and $5\times{}10^{12}\,{}M_\odot{}$). The total number of CRGs in the simulation,
from $z=0.1$ to $z=2$ (from $z=0.2$ to $z=1$) is 57 (23). \\
\footnotesize $^{\rm b}$ The redshift is observed for 8 galaxies in the L04 sample (see the Text for details). An estimated redshift
(L04) has been used for the 17 CRGs without redshift measurement. Note that the area of the survey is different from the one of
our simulation (see text for details).
}
\end{flushleft}
\label{tab_1}
\end{table}

Unfortunately, a direct redshift estimate has been provided only for 8 of the CRGs of the L04
sample\footnote{Redshift estimates are provided for 6 of the 25 galaxies of the LO4 sample,
in particular the estimates reported in Table 1 of LO4 refer to CRGs identified with number 2; 3; 5; 7; 10; 20. 
Two more  galaxies of that sample (CRGs labeled as 9 and 12) are members of galaxy clusters CL~0303+1706 
($z=0.6564$) and CL~1601+4253 ($z=0.5382$), respectively (see e.g. Dressler \& Gunn (1992) for the redshift determination).}.
LO4 proposed to estimate the redshift of CRGs  by adopting a 'standard' absolute $V$ magnitude for CRGs (Appleton \& Marston 1997).
\begin{figure}
\center{{
\psfig{figure=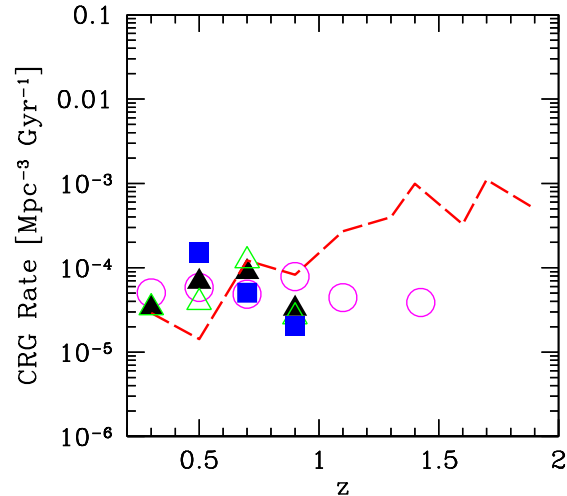,height=8cm}
}}
\caption{\label{fig:fig5} 
CRG formation rate per unit volume (comoving). Open triangles (case L04old, green in the online version): the sample of 25 observed CRGs in 
L04 with $z_{est}$ (the same as in figure~4 of L04). Filled squares (case L04$z_{obs}$, blue in the online version): the 8 observed CRGs with $z_{obs}$ 
(CRG number 2, 3, 5, 7, 9, 10, 12 and 20 of L04). Filled triangles (case L04new, black in the online version): the 8 observed CRGs with $z_{obs}$ and the remaining 17 observed CRGs 
with $z_{est}$. 
Open circles (magenta in the online version): CRG rate per unit volume derived from the E06 data.
Long dashed line (red in the online version): simulated CRGs. 
}
\end{figure}

Note that only the estimated redshifts ($z_{est}$) are shown in 
figure~4 of L04, even for those CRGs that have an observed redshift ($z_{obs}$). 
Also, the difference between $z_{est}$ and $z_{obs}$ is $\gtrsim{}$10 per cent, i.e. half of the bin width, for 4 of the 8 CRGs 
with  known redshift. Yet, for 3 of these 4 galaxies the observed redshift $z_{obs}$ is significantly smaller than 
the estimated value $z_{est}$ .

This discrepancy has important consequences in the estimate of the CRG formation rate. In Figure ~\ref{fig:fig5}
the CRG formation rate per unit volume\footnote{The total solid angle of the observations in L04 and in E06 was $\sim{}7.31\times{}10^{-5}$ sr and $\sim{}4.87\times{}10^{-5}$ sr, respectively, independent of redshift.} of the CRG sample observed by LO4 is shown as a function of redshift. For the open triangles (hereafter case 'L04old') of Figure~\ref{fig:fig5}, we assumed  the estimated redshift $z_{est}$
reported in LO4 for all the 25 CRGs (including the 8 CRGs with measured redshift $z_{obs}$). Note that the case L04old is the same as figure~4 of L04.
The formation rate of the 8 CRGs with observed redshift $z_{obs}$ is represented by the filled squares of Figure~\ref{fig:fig5} (hereafter 'L04$z_{obs}$').
Note that the ring galaxy formation rate of the sample of LO4 peaks at $z=0.7$ when $z_{est}$ is assumed (L04old),
whereas the peak shifts to $z=0.5$ when $z_{obs}$ is assumed instead (L04$z_{obs}$).

For completeness, we plot the rate of formation of 25 CRGs of LO4 when  $z_{obs}$ is assumed for 8 galaxies which have 
an observed redshift
and $z_{est}$ assumed for the remaining 17 CRGs (filled triangles, hereafter 'L04new'). 
The distribution peaks at $z\sim{}0.7$.
The open circles in Figure~(\ref{fig:fig5}) represent the CRG formation rate derived using the data reported by E06.
In this case all the redshifts are measured (either with spectroscopy or photometry).

We additionally plot in Figure~\ref{fig:fig5} the CRG formation rate per unit volume 
derived from cosmological simulations, as a function of redshift  
(dashed line). Note that the CRG formation rate is derived from Table~2 and is the same as shown in Figure ~\ref{fig:fig4}, but with a different normalization. The simulated CRG formation rate per unit volume approximately matches the data points in the redshift range $0.2\le z \le 0.8$.

We also note that in our simulations the CRG formation rate per unit volume substantially increases with redshift. As already noted for the merger rate,
this is mainly due to a substantial increase of the number density of progenitors present in merger tree at higher redshift.

At $z\sim{}1.1-1.5$ our simulated CRG formation rate is a factor of $\sim{}6-22$ higher than the one derived from E06 data. This might be due to several
factors. First, we assume that all progenitors with mass $10^{11}-5\times{}10^{12}\,{}M_\odot{}$ and experiencing interactions with small impact parameter are disc galaxies, but the morphology of the progenitor cannot be inferred from our dark matter only simulations. 
Second, the E06 sample might be considered incomplete at  $z > 1.1$.

Thus, new redshift measurements of moderately high-redshift CRGs will be extremely useful. The future survey 
zCOSMOS  (Lilly et al. 2007) 
will acquire spectra and redshifts of approximately 10,000 galaxies ($0<$ z $< 3$) in the COSMOS survey field  
and will derive 
for 10,000 galaxies between $0<$z$<1$ the mass, the morphology and the size. This will provide an excellent sample to compare with the rate of CRGs predicted from our models.

\subsection{Evolution of the Number of CRGs with Redshift}

Various theoretical models suggest that the galaxy merger rate  
per unit volume scales with the redshift as $\dot{n}\propto{}(1+z)^m$. 
The exact value of $m$ depends on the details of the adopted formalism, as well as 
on the cosmological parameters. Approaches based on the Press-Schechter formalism (Carlberg 1990a, 1990b) 
predict $m\sim{}2.5$, assuming $\Omega{}_{\rm M}=0.238$ from the WMAP3 constraints.

We showed in Figure~\ref{fig:fig4} that the CRG formation rate the is a good tracer of the merger rate. Hence, 
one might assume that also the CRG rate scales with $(1+z)^m$ (LO4). Thus, we derive the value of $m$ from the 
simulated CRGs. In particular,   
the expected number of CRGs ($N_{\rm CRG}(z_1,\,{}z_2)$), which form between $z_1$ and  $z_2$,  in a given volume, under the assumption that the density of CRGs scales as in  
equation~\ref{eq:eq1},
may be expressed by the following formula (see section 3.2 of L04).

\begin{equation}\label{eq:lavery}
\begin{array}{l}
N_{\rm CRG}(z_1,\,{}z_2)=n_{CRG,0}\,{}\left(\frac{c}{H_0}\right)^3\,{}\times{} \\
\hspace{2cm}\int_{z_1}^{z_2}(1+z)^m\,{}
\left[\int_0^z\frac{d\tilde{z}}{{\mathcal E}(\tilde{z})}\right]^2\,{}\frac{dz}{{\mathcal E}(z)}\,{}\Delta{}\Omega{}(z),\\
\end{array}
\end{equation}

where $c$ is the light speed, 
$n_{CRG,0}\sim{}5.4\times{}10^{-6}\,{}h^3\textrm{ Mpc}^{-3}$ 
is the current density of CRGs (Few \& Madore 1986), 
${\mathcal E}(z)=[(1+z)^3\Omega{}_M+\Omega{}_\Lambda{}]^{1/2}$  
and $\Delta{}\Omega{}(z)$ is the considered solid angle at a given redshift $z$.

In the simulations $\Delta{}\Omega{}(z)$ is the total solid angle, at a given $z$, occupied by our 
halo sample. 
$\Delta{}\Omega{}(z)$ has been calculated as the sum of the physical sizes of each halo in the simulation divided by the comoving distance at a given $z$, i.e.

\begin{equation}\label{eq:angle}
\Delta{}\Omega{}(z)\simeq{}8.6\times{}10^{-4}\,{}{\rm sr}\,{}\left(\frac{a}{0.769}\right)^{2}\,{}\left(\frac{\int_0^z\frac{dz}{{\mathcal E}(z)}}{0.283}\right)^{-2},
\end{equation}
where $a$ is the cosmological factor of expansion. 
The equation~(\ref{eq:angle}) is normalized to $z=0.3$.

Solving equation~(\ref{eq:lavery}) for the values of $\Delta{}\Omega{}(z)$ derived in equation~(\ref{eq:angle}) gives the expected number of CRGs which form in our simulations between $z_1$ and $z_2$, provided that the CRG density scales with $(1+z)^m$. 

\begin{table}
\begin{center}
\caption{Number of expected CRGs in our simulation, as a function of $m$, in the redshift range $z_1,\,{}z_2=(0.2,\,{}1.0)$ (central column) and $z_1,\,{}z_2=(0.2,\,{}2.0)$ (right column).}
\begin{tabular}{ccc}
\hline
\vspace{0.1cm}
$m$ & $N_{\rm CRG}$ & $N_{\rm CRG}$\\
& ($z_1=0.2,\,{}z_2=1.0$) &  ($z_1=0.2,\,{}z_2=2.0$)\\
\hline
1 &  6.7 & 10.0\\
2 & 10.4 & 18.5\\
3 & 16.5 & 36.4\\
4 & 26.7 & 76.5\\
5 & 44.0 & 170.1\\
6 & 73.8 & 397.1\\
\hline
\end{tabular}
\end{center}
\end{table}
We calculated equation~(\ref{eq:lavery}) for different values of 
$m$ and for $(z_1,\,{}z_2)=0.2,\,{}1.0$ and $(z_1,\,{}z_2)=0.2,\,{}2.0$. Results are reported in Table~3.
Our cosmological simulations show that 23 CRG
form between $z_1=0.2$ and $z_2=1.0$, leading to the value of $m\sim{}3.7$, which is lower than the one
obtained by L04 ($m\sim{}5$).
However, if we consider the simulated CRGs forming up to $z_2=2$, $m$ 
is slightly lower: 46 CRGs are expected to form if $m\sim{}3.3$.

If we adopt the model by Carlberg (1990a, 1990b), 
a value of $m=3-4$ is in disagreement with the cosmological parameters measured by WMAP3. 
In fact, equation~(\ref{eq:eq2}) implies $\Omega{}_{\rm M}\sim{}0.38-0.76$ for $m=3-4$, at odds with the estimate 
of WMAP3.

{\it Viceversa} if  $\Omega{}_{\rm M}=0.238$ is assumed, in agreement
with WMAP3 data, a value of $m\sim{}2.5$ is derived from equation~(\ref{eq:eq2}). For $m\sim{}2.5$ the predicted number 
of newly formed CRGs in simulations up to $z \sim 1$ is $\approx 13$ (see Table 3),
too low with respect to the observations of LO4.

\begin{figure}
\center{{
\psfig{figure=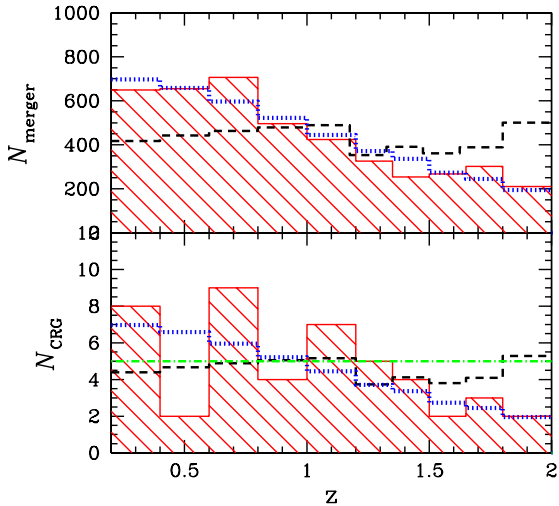,height=8cm}
}}
\caption{\label{fig:fig6} 
{\it Bottom panel:} number of newly formed CRGs as a function of redshift. Hatched histogram (red in the online version): simulated CRGs. Open histogram with dashed line: model expressed by equation~(\ref{eq:lavery}) for $m=3.3$ (see the Text and L04). Open histogram with dotted line (blue on the web): model expressed by equation~(\ref{eq:cons}) for $\gamma{}=50$, $m=2.4$ and $\beta{}=-2$ [see the Text and Conselice (2006)]. Open histogram with dot-dashed line (green on the web): flat distribution.
{\it Top panel:} total merger rate (hatched histogram, red on the web). Open histogram with dashed line: equation~(\ref{eq:lavery}) for $m=3.3$. Open histogram with dotted line (blue on the web): equation~(\ref{eq:cons}) for $\gamma{}=5000$, $m=2.4$ and $\beta{}=-2$.
}
\end{figure}

Furthermore, in the bottom panel of Figure~\ref{fig:fig6} we plot the evolution in redshift of the number of CRGs obtained in cosmological simulations    
(hatched histogram) compared with the number of CRGs  predicted from equation~(\ref{eq:lavery}) (open histogram with dashed line).
We note the different evolution of the two distributions with redshift: the former decreases, the latter increases with redshift. 
The null hypothesis probability that the 
two histograms are drawn from the same distribution is 0.19, corresponding to a non-reduced $\chi{}^2=12.4$. 
(with 10 data points, 1 parameter, and assuming Poissonian errors).

We  also compare the number of CRGs obtained in our cosmological simulations with the 
fitting function
proposed by Conselice (2006) 
[see equation (5) of Conselice (2006)]. The fitting formula by Conselice (2006), adapted for our simulated CRGs, can be written as

\begin{equation}\label{eq:cons}
N_{\rm CRG}(z)=\gamma{}\,{}(1+z)^m\,{}\exp{[\beta{}(1+z)]},
\end{equation}
where $\gamma{}$=50, $m$=2.4 and $\beta{}$=-2.0 best fit the simulated CRG distribution (between $z=0.2$ and $z=2.0$), reported in Figure ~\ref{fig:fig6}.

In  this case, the non-reduced $\chi{}^2$ is 7.5, leading to a null hypothesis probability equal to 0.38
(considering 10 data points, 3 parameters, and assuming Poissonian errors). 

However, the simulated CRG rate can be fit also by a flat distribution with $N_{\rm CRG}=5.0$ (non-reduced $\chi{}^2= 12.4$ and null hypothesis probability equal to 0.19, for 10 data points, 1 parameter and Poissonian errors).

In Section 3.3 we stressed that the CRG formation rate is a good tracer of the merger rate. For comparison, the top panel of Figure~\ref{fig:fig6} shows the evolution with redshift of the total merger rate (i.e. the merger rate of haloes with mass between $5\times{}10^{12}$ and $10^{14}\,{}M_\odot{}$ at $z=0$). 

In this Figure, the  total merger rate is compared with the Carlberg model (dashed line)  and with Conselice's formula (dotted line). 
Even in this case, the Carlberg model ($m=3.3$) does not match results from simulations, as it predicts an increase of the number of mergers with redshift. 
Instead, Conselice's formula is in better agreement with the simulated merger rate. 
In particular, the best-matching parameters for equation~(\ref{eq:cons}) are $\gamma{}$=5000, $m$=2.4 and $\beta{}$=-2.0. Then, Conselice's formula reproduces the evolution of the CRG formation rate as well as of the total merger rate.

\section{Summary}
We have used cosmological numerical simulations  to study the rate of mergers of haloes with mass $> 5\times{}10^{12}\,{}M_{\odot}$
and the CRG formation rate.   The large volume combined with the selection criterion used to identify  halo progenitors
allows us to quantify the cosmic merger fraction, the merger rate and the CRG formation rate among haloes. We have made comparisons between these and theoretical models and the latest available observational data.

Our main conclusions may be summarized as follows.
\begin{itemize}

\item The merger fraction of progenitors of the present day galaxies does not evolve strongly with the redshift between 0.2$\le$z$\le$2.
Predictions of the merger fraction and merger rates are in fair agreement with the current observational data, within the great 
uncertainties. This implies that the  observed decrease
of the cosmic star formation rate since z$\sim$1 is not tied to a disappearing population of major mergers, at least according
to our models. We calculate that the number of mergers a progenitor of a halo with mass $> 5 \times{} 10^{12}\,{}M_{\odot}$ will undergo 
from z$= 2$ to z$= 0.2$ is $N_m \sim 2$. We find  that there are still major mergers occurring at redshift lower than z$\sim$ 1, mainly due to the late assembly of large mass systems.

\item The formation rate of CRGs is a good tracer of the merger rate.

\item Assuming that the galaxy interaction rate per unit volume is proportional to $(1+z)^m$,  as suggested by previous models,
we derive $m=3-4$ from
our numerical simulations of the concordance cosmological model. However, the CRG formation rate as well as  the global (major+minor)  merger rate are best-matched by the formula $N_{\rm merger}=\gamma{}\,{}(1+z)^m\,{}\exp{[\beta{}(1+z)]}$ (Conselice 2006). 

\item The CRG formation rate inferred by simulations is in marginal agreement with the observed CRGs between $0.2\le$z$\le 1$. 
However, new redshift measurements 
are required to have a good statistical sample of CRGs. Future surveys like zCOSMOS will be able to provide insights on the incidence of 
galaxy mergers and CRGs rates, and will be extremely useful to test the rates predicted in the hierarchical universe.
\end{itemize}

\section*{Acknowledgments}
We would like to thank suggestions from the anonymous referee which have improved
this work. We are grateful to Shardha Jogee for kindly providing the GEMS data shown in Figure 1
and Figure 2 before publication and for a careful reading of the paper.
The numerical simulations were performed on zbox2 at University
of Zurich. We thank Jonathan Coles, Peter Englmaier, Joachim Stadel and Doug Potter for technical support, 
and we acknowledge Emanuele Ripamonti and George Lake for useful discussions. 
ED is supported by a EU Marie Curie Intra-European Fellowship
under contract MEIF-041569. MM acknowledges support from the Swiss
National Science Foundation, project number 200020-117969/1
(Computational Cosmology \&{} Astrophysics).

{}

\end{document}